\documentclass[pre,twocolumn,superscriptaddress]{revtex4}
\usepackage{epsfig}
\usepackage{subfigure}
\usepackage{amsmath}
\usepackage{amsfonts}

\begin{document}

\markboth{J.L. Belof, E.W. Lowe, R.W. Larsen, B. Space}
{Volume Determination of Globular Proteins by Molecular Dynamics}

\title{Volume Determination of Globular Proteins by Molecular Dynamics}
\date{\today}

\author{Jonathan L. Belof\footnote{
Present address: belof1@llnl.gov, Lawrence Livermore National Laboratory, 7000 East Ave., Livermore, CA 94550
}}
\address{Department of Chemistry, University of South Florida\\
4202 E. Fowler Ave., Tampa, FL, 33620, USA}

\author{Edward. W. Lowe}
\address{Center for Structural Biology, Vanderbilt University\\
465 21st Ave. South, Nashville, TN 37212, USA}

\author{Randy W. Larsen}
\address{Department of Chemistry, University of South Florida\\
4202 E. Fowler Ave., Tampa, FL, 33620, USA}

\author{Brian Space}
\address{Department of Chemistry, University of South Florida\\
4202 E. Fowler Ave., Tampa, FL, 33620, USA}

\begin{abstract}
Molecular dynamics simulations of myoglobin and aspartate aminotransferase, with explicit solvent, are shown to accurately reproduce the experimentally measured molar volumes.  Single amino-acid substitution at VAL39 of aspartate aminotransferase is known to produce large volumetric changes in the enzyme, and this effect is demonstrated in simulation as well.  This molecular dynamics approach, while more computationally expensive that extant computational methods of determining the apparent volume of biological systems, is quite feasible with modern computer hardware and is shown to yield accurate volumetric data with as little as several nanoseconds of dynamics.

\end{abstract}

\maketitle

\section{Introduction} 
\label{sec:intro}

Experimental techniques for determining the partial specific volume and partial specific adiabatic compressibility of proteins in solution have provided key insight into structural and catalytic events.  The application of these methods has resulted in a broad base of knowledge about solvation effects, ligand binding and dissociation, the influence of protein domains on catalytic events, and protein folding pathways.\cite{chu,gekko1,gekko2,jager}

Theoretical methods offer the potential to link specific events to the thermodynamic observables that experimentalists measure in the course of their research.  In particular, molecular dynamics (MD) can provide a powerful approach to correlating the molecular trajectories of proteins that may give rise to an experimentally measured molecular volume or change thereof.

The state-of-the-art method to date for determining the apparent volume of a protein is the method of ``accessible surface area'',\cite{chalikian} a method that employs a spherically-accessible surface integration of the protein’s crystal structure.  This method is attractive in that it is not computationally demanding (and is thus accessible to the typical computational equipment of an experimental lab), however it calculates volumes that can be significantly different from those measured in solution.

Prior research on the volume of small molecules has been performed by our group\cite{devane,ridley} and the methodology presented here is an application of those techniques to protein systems.  This work demonstrates the validity of the MD approach toward determining the apparent molecular volume of globular proteins.

\begin{figure}[htp]
\includegraphics[width=3.3 in]{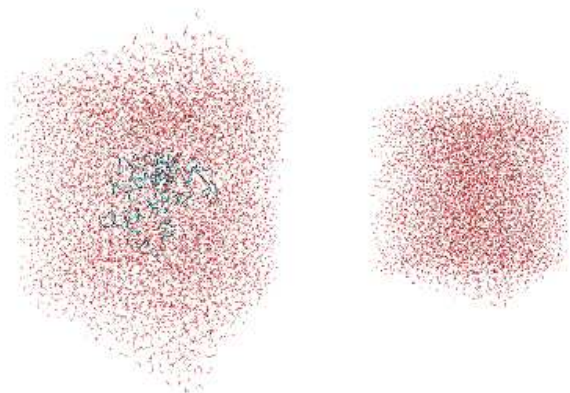}
\caption{A periodic cell containing horse-heart myoglobin solvated with water molecules, in comparison with a system containing only the bulk TIP3P water (both simulations cells contain 10,796 water molecules).  The difference in the volume between each cell under NPT dynamics results in the apparent molecular volume of the myoglobin.}
\label{fig:box}
\end{figure}

\section{Methods} 
\label{sec:methods}

The apparent molecular volume of a protein is calculated as:

\begin{eqnarray}
V_p = V_{p+w} - V_w
\end{eqnarray}

The NAMD\cite{phillips} molecular dynamics package uses a Langevin-Hoover hybrid method where a piston is coupled to the equations of motion for a particle in the isothermal-isobaric (NPT)ensemble.

A simulated annealing algorithm was implemented to perform a global energy minimization after making single amino-acid substitutions.  The protein and water system is brought to a higher energy state and allowed to randomly walk through phase space as the system is cooled over a specific temperature schedule.

Post-simulation analysis consisted of calculating the correlation time of the volume signal using a block-averaging method in order to calculate an unbiased error in the volume, as well as structural analysis of the equilibrium structures (RMSD, Ramachandran plots) versus their crystal structures.

\section{Results and Discussion}
\label{sec:results_and_discussion}

\subsection{Horse-heart myoglobin}
\label{subsec:hh_myoglobin}

Preparatory simulation stages (NAMD\cite{phillips}) consisted of constructing the solvated protein system and local energy minimization, followed by heating and equilibration steps.  Initial myoglobin structural coordinates were obtained from the RCSB protein data bank (crystal structure entry 1DWR4), solvated with 10,796 TIP3P water molecules and then minimized by the method of conjugate gradients.  The protein was not mutated in any way.  The system was then heated to 300 K over a period of 1.2 ns, followed by equilibration in the NPT ensemble for 0.2 ns.

A production NPT (300 K/1 atm/2.0 fs timestep) run of 10.0 ns resulted in an average volume for the protein-water system of 341,875 $\textrm{\AA}^3$; simulating the bulk water alone over a trajectory of equal time duration yielded an average system volume of 319,775 $\textrm{\AA}^3$.  The difference of these values gives an apparent molecular volume of wild-type myoglobin corresponding to 22,100 $\textrm{\AA}^3$ 0.747 cm$^{3}$ / g.  The correlation times of both volume signals were determined by block-averaging and the signals were then uncorrelated to provide an unbiased volume error of $\pm$ 0.001 cm$^{3}$ / g.  This computationally determined apparent volume of 0.747 $\pm$ 0.001 cm$^{3}$ / g agrees precisely with experimentally reported sound velocity measurements\cite{chu} and is within the experimental error of that study.  The equilibrium protein structure in solution was aligned and compared with the crystal structure of myoglobin and while the overall RMSD was minimal, a local region of amino acids near GLY80 was found to be displaced by 6.0 $\textrm{\AA}$ due to solvation.

\begin{figure}[htp]
\includegraphics[width=3.3 in]{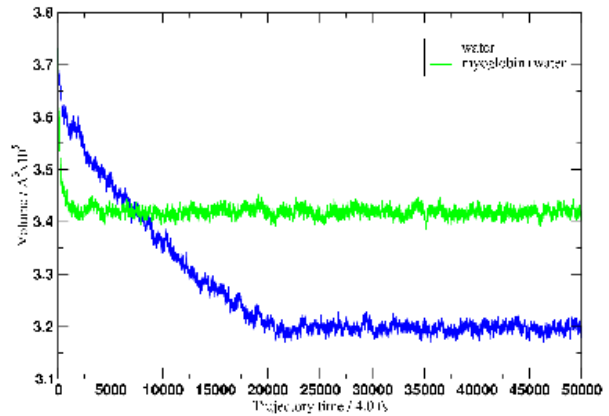}
\caption{Volume signals for the solvated myoglobin system vs. the bulk water.  After initial equilibration the volumes fluctuate about their average values for the 10 ns trajectory – the difference of which gives an apparent molecular volume of 0.747 cm$^{3}$ / g.}
\label{fig:graph}
\end{figure}

\subsection{Aspartate aminotransferase}
\label{subsec:aspat}

The halozyme of E. coli aspartate aminotransferase (RCSB entry 1ASM)\cite{gekko1,gekko2} a large dimer consisting of identical 404 amino acid subunits complete with LYS258-bound pyridoxal-5'-phosphate cofactors, was simulated in the NPT ensemble with 58,361 water molecules.  Both native AspAT and it's VAL39 mutant were simulated in order to compare molar volumes. An average apparent volume of 0.733 cm$^{3}$ / g was calculated at the end of a 0.5 ns run, a result that is in good agreement with the experimentally measured value of 0.731 cm$^{3}$ / g. Work is in progress to obtain a longer timescale trajectory and calculate the associated volumetric error.

A series of single-point mutations at VAL39 have been investigated experimentally to determine their effect on compressibility and volume. VAL39 was chosen due to its proximity to the binding site and was theorized to serve as a gating amino acid influencing substrate specificity. A positive linear correlation between adiabatic compressibility and apparent volume when bulky side chains were introduced suggested that these effects were due to the increased flexibility of the protein and an increase in cavity size caused by these mutations. The largest effect on protein volume was observed for the V39G variant causing a shift from 0.731 cm$^{3}$ / g to 0.696 cm$^{3}$ / g. This single-point mutation induced a large conformational change in the dimer\cite{jager} and an associated change in the apparent volume by -0.035 cm$^{3}$ / g, one of the largest molecular volume changes observed due to a single residue mutation relative to the native protein. Since the crystallographic structure of the V39G mutant has not yet been elucidated, manual alteration of the VAL39 side chain of the 1ASM crystal structure was performed to give the V39G initial configuration. A simulated annealing algorithm was developed and performed on the mutant to help facilitate the adoption of its new equilibrium conformation prior to performing production NPT runs.

\begin{figure}[htp]
\includegraphics[width=3.3 in]{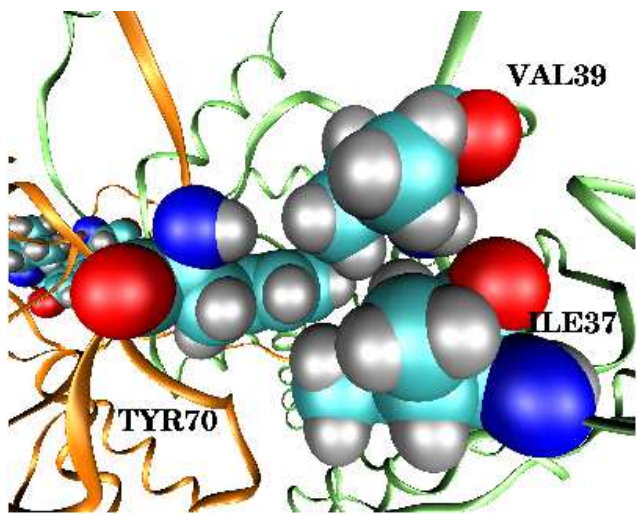}
\caption{Van der Waals representation showing the steric effects of the valine (V39), tyrosine (Y70 on the second subunit of the dimer) and isoleucine (I37) residues of interest for the V39G mutant active site.  Substitution of the valine for a glycine reduces the steric interactions leading to a large scale conformational change as well as a drastic change in catalytic activity.  The same triad can be seen in the distance to the left on the opposite side of the dimer.}
\label{fig:aspat}
\end{figure}

\section{Conclusions}
\label{sec:conclusions}

The application of molecular dynamics for studying the volume of globular proteins can accurately model experimental data.  The methods presented can offer insight into structural protein studies since conformational changes can be examined and correlated with experimentally observed volume changes in solution.  The volumetric contribution of various regions of a protein (including the more difficult case of a single-point mutation) can be elucidated through use of this practical and consistent methodology.

Work is currently in progress to resolve the separate Coulombic and Van der Waals contributions to the apparent volume.  Hybrid Monte Carlo (HMC) methods are also being developed to more efficiently explore the phase space of folding intermediates and to allow the use of additional potential energy terms.

\section*{Acknowledgments}
\label{sec:acknowledgments}

Computations were performed at the USF Research Computing Center where NSF-funded computational resources (under Grant No. CHE-0722887) were greatly appreciated.  The authors acknowledge funding from the National Science Foundation (Grant No. CHE-0312834).  The authors also thank the Space Foundation (Basic and Applied Research) for partial support.

\end{document}